\documentclass{aa}  
\usepackage{array}
\usepackage{textcomp, gensymb}
\usepackage{comment}
\usepackage{xcolor}
\usepackage{footmisc}
\usepackage{tablefootnote}
\usepackage{float}
\usepackage{commath}
\usepackage{sidecap}
\usepackage{caption}
\usepackage{natbib}
\usepackage{eht}
\usepackage{calrsfs}
\DeclareMathAlphabet{\pazocal}{OMS}{zplm}{m}{n}

\usepackage{graphicx}
\usepackage{float}
\usepackage{txfonts}
\usepackage{hyperref}
\hypersetup{colorlinks,linkcolor={blue},citecolor={blue},urlcolor={blue}}
\begin{document} 

   \title{Modeling MAD \sgra flares and their dependence on spin} 
%
   \author{Eleni Antonopoulou\inst{1}\fnmsep\inst{2}, Antonios Nathanail\inst{1}, Yosuke Mizuno\inst{3,4,5},  Foivos Zanias\inst{1}}
   \authorrunning{E. Antonopoulou, A. Nathanail. Y. Mizuno, F. Zanias}
   \institute{Research Center for Astronomy and Applied Mathematics, Academy of Athens, Soranou Efesiou 4, 115 27 Athens, Greece
   \and
   Department of Physics, National and Kapodistrian University of Athens, University Campus, GR 15784 Zografos, Greece
    \and
    Tsung-Dao Lee Institute, Shanghai Jiao Tong University, Shanghai, 201210, People’s Republic of China
    \and
    School of Physics and Astronomy, Shanghai Jiao Tong University, Shanghai, 200240, People’s Republic of China
    \and
    Key Laboratory for Particle Astrophysics and Cosmology (MOE) and Shanghai Key Laboratory for Particle Physics and Cosmology, Shanghai Jiao Tong University, Shanghai 200240, People’s Republic of China
             }
   \date{Received Month Date, 2026; accepted Month Date, 2026}
  \abstract
   {Magnetically arrested disks (MADs) represent one of the most promising theoretical frameworks to describe gas accretion and flare emission in the vicinity of a supermassive black hole. 
   State-of-the-art numerical simulations demonstrate great similarities between the MAD regime and the observational signature of Sagittarius A*. 
   In addition, recent work shows that the quasi-periodic flux eruption events seen in MAD simulations are suitable candidates for reproducing the observed flaring activity in the Galactic center.
   } 
   {This study aims to place constraints on the spin of \sgra from recent flare observations. 
   It delves deeper into the MAD state, analyzes the dynamics of flux eruptions, investigates the three dimensional morphology of magnetized flux tubes, and explores their relation to the spin of the black hole. 
   This research models the flare kinematics in the vicinity of a supermassive black hole, evaluates their dependence on black hole spin, and assesses the correlation between the spin of the black hole and the astrometric observations in the Galactic center. 
   }
    {\cite{Antonopoulou2025} modeled the flare kinematics in a MAD simulation featuring a counterclockwise accretion disk and a prograde black hole spin of $a=+15/16$.
    Our study extends this work to a range of black hole spins, including retrograde simulations and the non-spinning Schwarzschild case. 
    The methodology is as follows:
    a prominent flux eruption is identified in the simulation, and the evolution of the active region on the equatorial plane is tracked throughout the event.
    The flare model consists of a spherical hot spot ejected along the newly generated flux tube with a relativistic velocity.
    General relativistic radiative transfer calculations are performed and the resulting flare trajectory is compared to the observations of \cite{GRAVITY_2018}.
    }
    {In MAD simulations, the flux tube wraps around the jet funnel in the opposite direction to the black hole's spin.
    This dependence is imprinted directly on the flare kinematics, as the ejected hot spot trajectory traces the underlying flux tube morphology and provides a strong correlation between the spin of the black and the flare trajectory on the sky.
    The present analysis favors Kerr black holes, with a preference towards larger spin magnitude. 
    Flare trajectories exhibit a stronger clockwise bend for increasing spin magnitude that is in good agreement with the observed flare kinematics in the Galactic center. 
    Although prograde simulations show a consistent clockwise motion in the sky regardless of disk rotation, retrograde models require a clockwise accretion disk to account for the observations of July 22, 2018.
    }
    {A maximally rotating Kerr black hole is the most promising candidate for reproducing the observed flare trajectory in the vicinity of \sgra, illustrating a pass faction surpassing $90\%$ among the investigated flare models. 
    In particular, determining spin orientation within this model is linked to the rotation of the accretion disk.
    Assuming a counterclockwise accretion disk, a prograde black hole with $a\geq1/2$ is able to reproduce the observed flare kinematics in the Galactic center.
    Conversely, assuming a clockwise rotation for the disk, the observations support both prograde and retrograde Kerr black holes with a large spin.
    }
   \maketitle
%
\section{Introduction}
The supermassive black hole in the Galactic Center, hereafter \sgra, has been the target of intensive research and several monitoring campaigns throughout the years. 
One of its key properties is the frequent flare emission in the near-infrared (NIR) \citep{genzel2003,Ghez_2004,witzel2018}, mid-infrared \citep{vonFellenberg_2025,Michail_2026}, and X-ray regime \citep{baganoff2001,Neilsen2013,Ponti2015} roughly once per day. 
Every X-ray flare to date has had a simultaneous NIR counterpart, although the reverse is not true; most NIR flares do not show a detectable X-ray counterpart \citep{genzel2003,eckart2006,boyce2019}.
Interestingly, NIR flaring emission is highly variable and has been linked to a clockwise flux centroid motion close to the supermassive black hole \citep{GRAVITY_2018,GRAVITY_2020}.
Several of these "hot spots" have been observed in the past years, demonstrating a relativistic clockwise pattern in the sky, a few gravitational radii from \sgra, and a matching evolution in polarization angle over an hourly cadence \citep{GRAVITY_2020,GRAVITY_2020_polarimetry,GRAVITY_2023}. 

Since their first detection, the astrometric and polarimetric properties of these flares have been the focus of several theoretical and numerical investigations \citep{Tursunov_2020,Ball_2021,ElMellah2022,Vos_2022,vincent_2023,yfantis_2023,Lin_2023,Aimar_2023,mahdi_2023,Lin_2023,Zhdankin2023,Galishnikova2023,ElMellah2023,Lin_2024,Huang_2024,kocherlakota_2024,Lin_2024,Ricarte2025,Aimar2026}. 
The flaring events are usually attributed to prominent hot spot emission over a subdominant background, a few gravitational radii from the black hole.
The prominent flux centroids move at relativistic velocities and show good agreement with an underlying super-Keplerian orbital frequency in the accretion flow \citep{Matsumoto_2020,Antonopoulou_2024}, although they can also be modeled with a sub-Keplerian velocity \citep{Ball_2021,Antonopoulou2025}.
In addition, they consistently trace clockwise trajectories in the sky that favor more face-on observation angles for the source \citep{GRAVITY_2018,Antonopoulou2025}.
Most recently, evidence of multiple hot spots orbiting \sgra has been found in the $340\,$GHz variability of the source \citep{Yanagisawa_2026}. 

State-of-the-art numerical simulations have highlighted magnetically arrested disk (MAD) models as suitable candidates for describing gas accretion in the vicinity of a supermassive black hole \citep{Bisnovatyi_Ruzmaikin_1974,Narayan_2003,Mizuno2021,CruzOsorio_2022,Fromm_2022,Scepi2022}.
MAD models are characterized by high values of magnetic flux at the event horizon and show a quasi-periodic cycle of flux eruption events \citep{Dexter_2020,Porth_2021,Ripperda_2022}.
Magnetic flux accretes onto the black hole until it reaches the saturation value ($\phi_{crit}\sim15$ \cite{tchekhovskoy2012,porth2019}) and becomes strong enough to balance the ram pressure of the disk \citep{Igumenshchev_2003,Igumenshchev_2008,Tchekhovskoy_2011}.
During these events, accretion of matter onto the black hole is significantly halted and can only proceed via means of the magnetic Rayleigh-Taylor instability \citep{Papadopoulos_2018,Loules2026}.

In addition, current sheets of alternating polarity are generated on the disk's equatorial plane, powering active magnetic reconnection sites \citep{Kagan2015,Ball2018,Comisso2019,Ripperda_2020,Ripperda_2022,Imbrogno2025,Meringolo2026} and creating chains of magnetized plasmoids that are then expelled into the disk \citep{Loureiro_2007,Moscibr_2007, Fermo_2010, Uzdensky_2010, Huang_2012, Loureiro_2012, Takamoto_2013, Nathanail_2020, Chatterjee_2021,Sukova_2021,Nathanail_2022,Lin_2023,Jia2023,Dimitropoulos_2024,Jiang2024,Xia2026}.   
Flux eruptions are a characteristic byproduct of the MAD accretion state and occur in a repeated and quasi-periodic manner within the corresponding simulations.
They are characterized by a sudden drop in magnetic flux on the event horizon ($\phi_{BH}$) and the overall image intensity \citep{Jia2023,Ricarte2026}.

\cite{EHTV2022} determined a promising cluster of MAD models, featuring a prograde black hole spin and a face-on observation angle, from horizon-scale and multi-frequency observations of \sgra.
Motivated by these results, we employed a MAD configuration surrounding a maximally rotating Kerr black hole to model the flare kinematics following a prominent flux eruption \citep{Antonopoulou2025}. 
We modeled a relativistic hot spot ejected along a magnetized flux tube, showed that it can balance the counterclockwise drag of the accretion disk, and demonstrate a clockwise motion in the sky that is in good agreement with the observed flare kinematics in the Galactic center \citep{GRAVITY_2018}. 
This study extends the work of \cite{Antonopoulou2025} by analyzing a suite of three dimensional GRMHD simulations in the MAD regime, ranging from prograde to retrograde black hole spin.
Utilizing the methodology developed, this study investigates the morphology of magnetized flux tubes generated during a flux eruption event and models the motion of hot spots assumed to be ejected from the equatorial plane.
We explore the dependence of the flux tube morphology and the flare kinematics on black hole spin, and search for a direct correlation with the observations of 2018 July 22 in the vicinity of \sgra. 

This paper is organized as follows: 
Section~\ref{Sec:Methodology} gives an overview of the methodology used to identify promising flux eruption events, track the active regions on the equatorial plane, and model flares in MAD simulations.  
Section~\ref{Sec:Results} depicts the flux tube morphology and the resulting flare kinematics, focusing on their relation to black hole spin. 
Section~\ref{Sec:Conclusions} presents the implications of this study on deriving black hole spin from NIR flare observations, while section \ref{sec:Discussion} discusses future avenues to be explored.
The numerical setup for the GRMHD simulations employed in this work is given in Appendix \ref{Append:Numerical_Setup} and the evolution of a magnetized flux tube during a flux eruption is discussed in Appendix \ref{Apped:Fluxtube_evol}.

\section{Methodology} \label{Sec:Methodology}
This study investigates a set of 3D MAD simulations performed with the Black Hole Accretion Code \citep[hereafter BHAC;][]{Porth_2017,Olivares_2019}.
Although our previous work employed a maximally rotating Kerr black hole with $a_{BH}=+15/16$, the current investigation expands this parameter space to explore the effect of black hole spin on flare dynamics.
We consider four distinct configurations: a moderately prograde case ($a_{BH} = +1/2$), a non-spinning Schwarzschild case ($a_{BH} = 0$), and two retrograde cases ($a_{BH} = -1/2$ and $a_{BH} = -15/16$).
All simulations feature a counterclockwise accretion disk, consistent with \cite{Antonopoulou2025}. 
Details on the numerical setup are given in Appendix \ref{Append:Numerical_Setup}.

\begin{figure}
\centering
\includegraphics[width=0.5\textwidth]{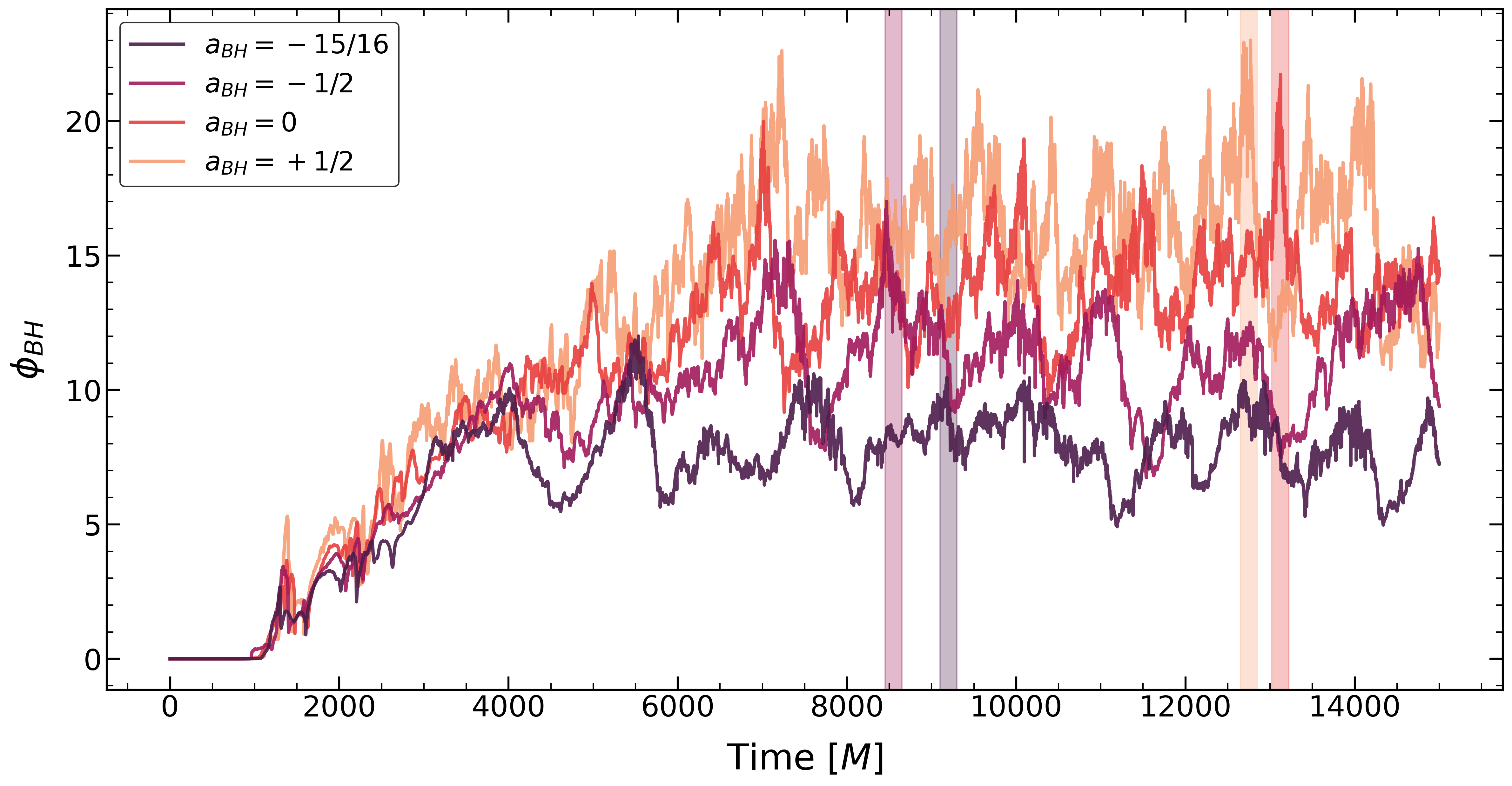}
\caption{Time-series of the normalized magnetic flux on the event horizon for the MAD simulations considered in this work. The investigated window of $\sim200\,$M during the selected flux eruption event is highlighted in color.}
\label{Fig:Mdot}
\end{figure}

The first step in our analysis consists of examining the time series of the mass accretion rate and magnetic flux over the surface of the event horizon for each simulation. 
In the context of a GRMHD simulation, these quantities can be defined as follows
\begin{align}
    \dot{M} &\equiv -\iint \rho u^r \sqrt{-g}\,d\theta\,d\phi \,, \\
    \Phi_{BH} &\equiv \dfrac{1}{2}\iint \,\abs{B^r}\sqrt{-g}\,d\theta\,d\phi \,,
\end{align}
where $\rho, u^r, g$ and $B^r$ are the mass density, radial four-velocity, metric determinant, and radial component of the magnetic field as seen by local fiducial observers, respectively.
The normalized magnetic flux on the horizon is then given by
\begin{equation}
    \phi_{BH}\equiv \sqrt{4\pi}\:\dfrac{\Phi_{BH}}{\sqrt{\dot{M}}} \,.
\end{equation} 
Fig.~\ref{Fig:Mdot} shows the time series of the dimensionless normalized magnetic flux for the investigated set of simulations.
After $\phi_{BH}$ saturates and the simulation reaches a quasi-stationary MAD state ($\phi_{BH}\sim 15$, see \cite{tchekhovskoy2012,porth2019}), a prominent flux eruption event may be identified as a characteristic "dip" in the time series; a clear signature of magnetic flux being expelled from the horizon.
For each simulation, we select a representative flux eruption and analyze a temporal window of approximately $200\,$M encompassing the event, as illustrated in Fig.~\ref{Fig:Mdot}. 

Flux eruption events are characterized by strong magnetic reconnection within the disk's equatorial plane \citep{Ripperda_2020,Ripperda_2022}. 
This process accelerates particle populations into forming energetic hot spots spanning a few gravitational radii close to the supermassive black hole \citep{Hakobyan_2023,Zhdankin2023,Hakobyan_2025}. 
This study considers spherical hot spots with a constant emission radius of $1\,$M \citep{GRAVITY_2018,Nathanail_2020,Vos_2022,Yfantis2026}$-$the effect of shearing throughout the trajectory is neglected$-$.
We assume that these hot spots are ejected from the disk plane and travel along magnetized flux tubes at a fraction of the speed of light \citep{Nathanail_2020,Vos_2022,Lin_2024}. 
General relativistic radiative transfer (GRRT) calculations are performed in Boyer-Lindquist coordinates with Black Hole IMaging \citep{Antonopoulou_2024}.
A face-on inclination is employed, motivated by the observed flare kinematics of July 22, 2018 \citep{Antonopoulou2025}.
Hot spot emission is modeled by a power-law electron distribution \citep{Pandya_2016}, while synchrotron absorption is neglected as a first-order approximation.
For a more detailed discussion of the modeling process, we refer the reader to \cite{Antonopoulou2025}.

\begin{figure*}[h]
\centering
\includegraphics[width=0.8\textwidth]{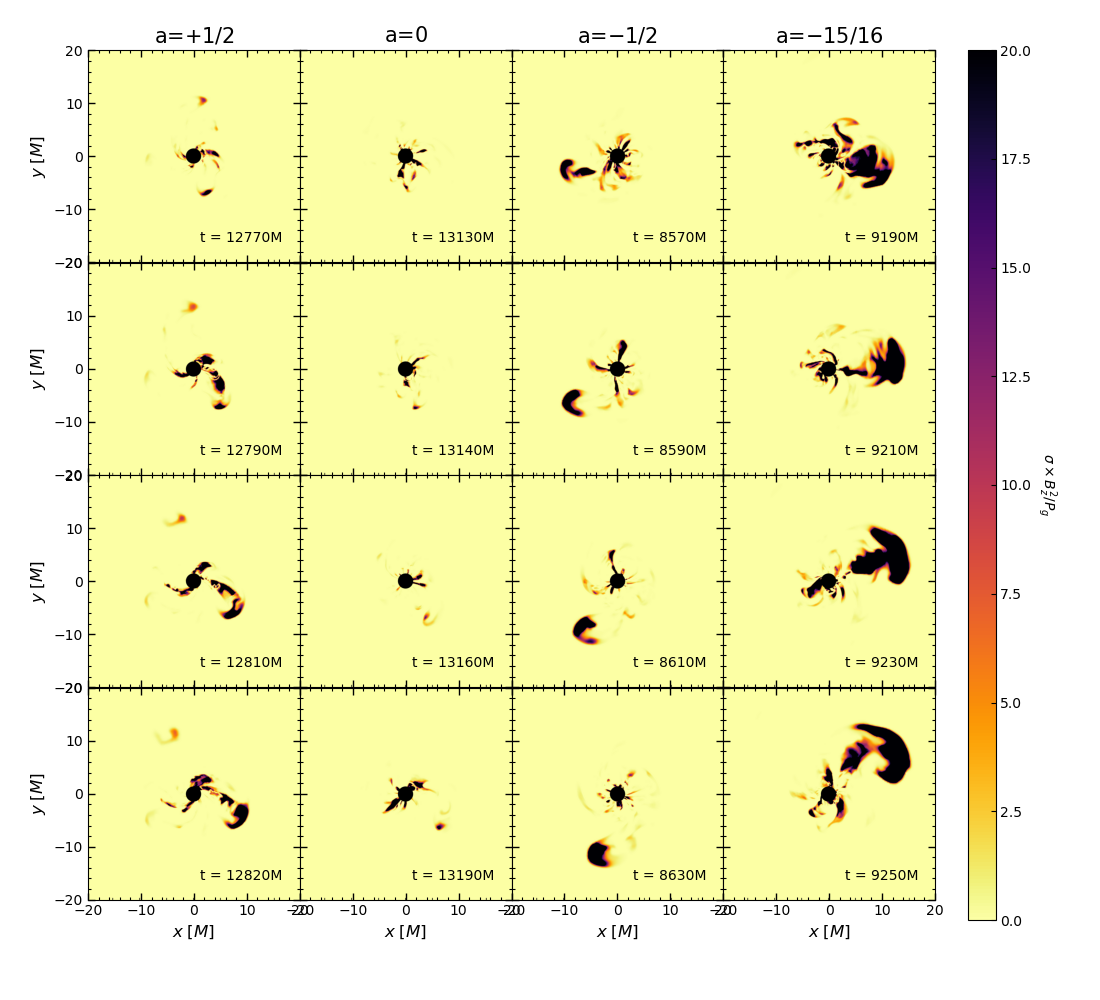}
\caption{The active region on the equatorial plane of the disk following a prominent flux eruption. The color corresponds to the product of the magnetization times the vertical magnetic field strength, divided by the gas pressure ($\sigma \times B_z^2/P_g$). From left to right: prograde ($a=+1/2$), nonspinning ($a=0)$, and retrograde ($a=-1/2$ and $a=-15/16$) simulations. The evolution of the active region in each simulation is shown from top to bottom. The corresponding timestamp (in code units) is denoted in the legend. 
}
\label{Fig:equator_sigmaB2}
\end{figure*}

In MAD simulations, flux tubes are dominated by large-scale vertical magnetic fields and can be traced back to their highly magnetized foot-points on the equatorial plane \citep{Porth_2021,Antonopoulou2025}.
Following the methodology of \cite{Antonopoulou2025}, we search for active regions on the equatorial plane, characterized by high magnetization, $\sigma\geq0.4$, and large vertical magnetic field strength, $B_z^2/P_g\geq0.8$.
The product of these two quantities is an ideal metric to differentiate between newly generated flux tubes, powered by recent flux eruptions, and stagnated magnetic flux from previous events that remains within the disk and moves passively with the accretion flow.
Fig.~\ref{Fig:equator_sigmaB2} illustrates the evolution of these active regions in the equatorial plane for the MAD simulations investigated.
These regions demonstrate both an orbital motion due to the accretion disk's rotation and an outward motion toward larger radii due to the flux eruption.

We recognize a distinct behavior between the prograde and retrograde simulations in Fig.~\ref{Fig:equator_sigmaB2}.
In retrograde simulations, the size of the active region is significantly larger, spanning up to ten gravitational radii in length in the maximally rotating case.
On the contrary, both the moderately spinning prograde case shown here and the maximally rotating case addressed in \cite{Antonopoulou2025} demonstrate a smaller and thinner active region.
Furthermore, this distinction is more pronounced for increasing black hole spin.
The active region deforms and extends in size as time passes; however, it becomes much thinner in width in prograde simulations, reducing nearly to a thin strip for $a=+15/16$ (see Fig.~1 in \cite{Antonopoulou2025}).
The effect is even more drastic in the non-spinning case, where the active sites are mostly contained close to the black hole and are much smaller in size.
As shown in Fig.~\ref{Fig:equator_sigmaB2}, the active region is the size of a small blob on the equatorial plane even after a prominent flux eruption event.

By isolating these active regions, one can accurately trace the foot-points of the magnetized flux tubes following the flux eruption.
As the foot-points move counterclockwise with the accretion flow, the flux tube experiences bending in the opposite direction due to rotation.
Its three-dimensional morphology stabilizes shortly after, showing minimal changes in the overall structure (see Appendix~\ref{Apped:Fluxtube_evol}).
Thus, a fixed flux tube shape is employed for each simulation, imported directly from the 3D datasets 
a few time units following the onset of the flux eruption (see bottom row panels in Fig.~\ref{Fig:equator_sigmaB2}).
We model the motion of hot spots, assuming that they are ejected along the magnetized flux tube at a fraction of the speed of light, and taking into account the motion of the active region on the equatorial plane. 

\begin{figure*}[htbp] 
    \centering
    \begin{minipage}{0.65\textwidth}
        \centering
        \includegraphics[width=0.45\linewidth]{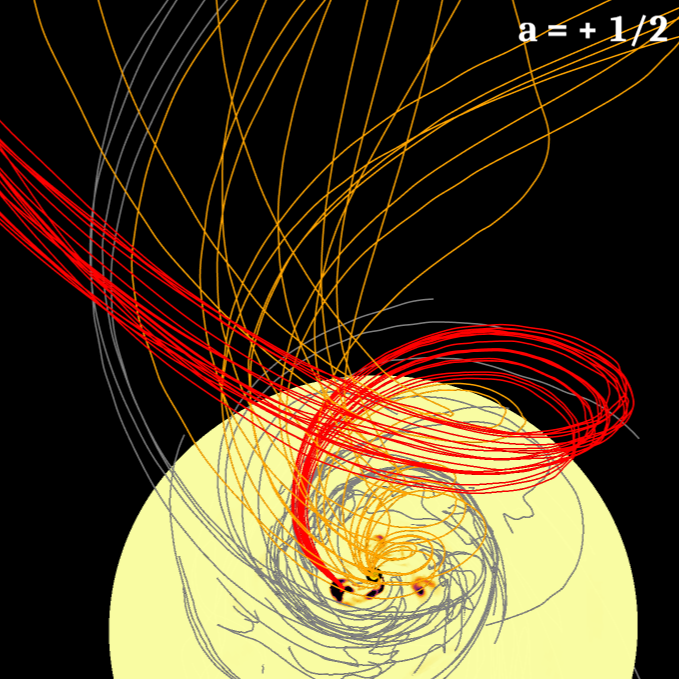}\hfill
        \includegraphics[width=0.45\linewidth]{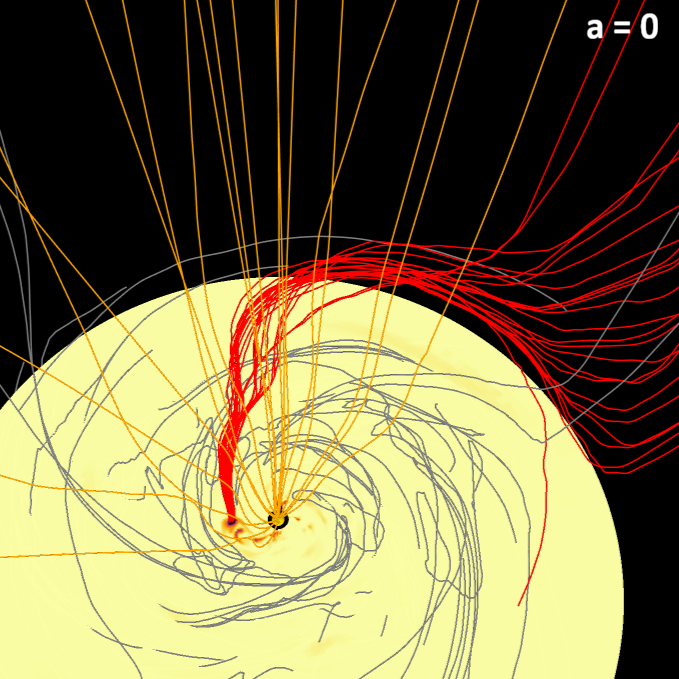}     
        \vspace{1cm} 
        \includegraphics[width=0.45\linewidth]{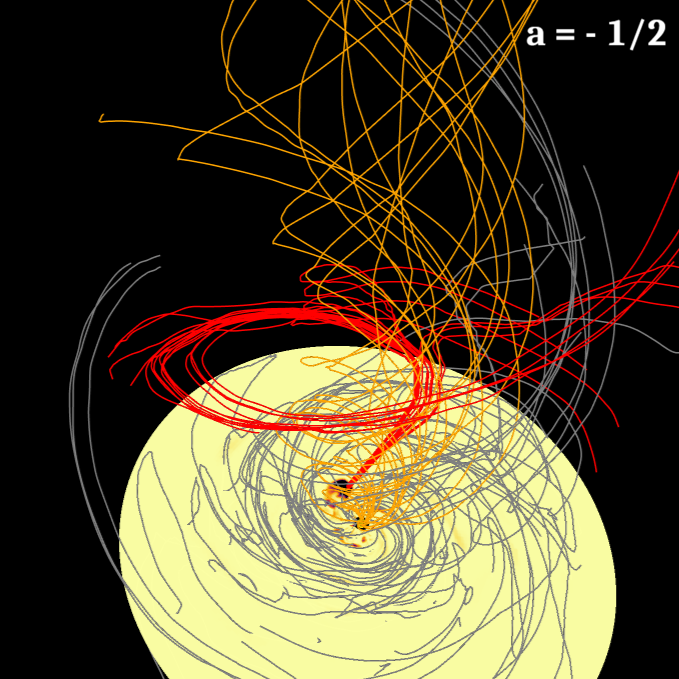}\hfill
        \includegraphics[width=0.45\linewidth]{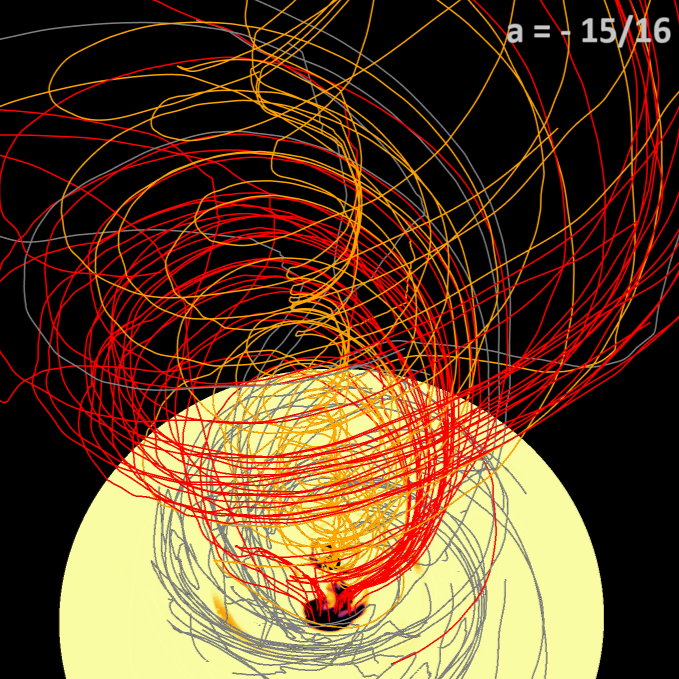}
    \end{minipage}\hfill
    \begin{minipage}{0.3\textwidth}
        \caption{Three-dimensional representation of the magnetic field topology during a flux eruption in the investigated MAD simulations. The color highlights the active region on the equatorial plane of the disk, where newly generated flux bundles emerge (red lines). The red field lines correspond to the most energetic part of the flux bundle. The gray lines represent the magnetic field lines of the accretion disk. The orange lines show the magnetized funnel field lines, originating from the black hole. 
        Illustrated from left to right: \textit{Top}: Prograde spin ($a=+1/2$) and nonspinning ($a=0$) simulations, \textit{Bottom}: Retrograde simulations ($a=-1/2$ and $a=-15/16$, respectively).}
        \label{Fig:flux_tube}
    \end{minipage}
    
\end{figure*}

\section{Results} \label{Sec:Results}
This section presents the results of applying the methodology outlined in section \ref{Sec:Methodology} in our set of MAD simulations.
First, we focus on the three-dimensional morphology of the magnetized flux tube generated during a prominent flux eruption event (section~\ref{Sec:Flux_Tube}).
Then, we illustrate the flare models for each simulation and compare them with the observations of \cite{GRAVITY_2018} (section~\ref{Sec:Flares}). 
In what follows, we discuss the dependence of the flux tube characteristics and flare kinematics on black hole spin and contrast our results with the maximally rotating case investigated in \cite{Antonopoulou2025}. 
\subsection{Flux tube morphology} \label{Sec:Flux_Tube}
Fig.~\ref{Fig:flux_tube} depicts the three-dimensional morphology of the magnetized flux tube for the investigated MAD simulations.
In all cases, the flux tube foot-points rest within the active region on the equatorial plane, as illustrated in the bottom panels of Fig.~\ref{Fig:equator_sigmaB2}.
We discover a clear distinction between the prograde and retrograde simulations.
Specifically, the prograde case with a moderate spin of $a=+1/2$ shows a helical flux tube winding up in a clockwise manner due to the accretion disk rotation.
The illustrated flux tube (depicted in red in Fig.~\ref{Fig:flux_tube}) demonstrates a structured magnetic field and a clear clockwise bend, continuously rising in height and wrapping around the magnetized jet funnel.  
In addition, the flux tube morphology in the prograde simulation is very similar to the maximally rotating case\footnote{We refer the reader to \cite{Nalewajko_2026} for a detailed analysis of the bending and wrapping of flux tube magnetic field lines around the jet funnel.} presented in \cite{Antonopoulou2025}.
The three-dimensional flux tube is clearly distinguishable from the magnetic field topology outside the active region, which features more tangled magnetic field lines that are not typically able to reach such large heights above the equatorial plane. 

The non-spinning simulation depicted in Fig.~\ref{Fig:flux_tube} shows a very similar behavior to the prograde case.
The illustrated flux tube morphology features a consistent clockwise bend as it wraps around the magnetized jet funnel.
Since the active region is noticeably smaller in size (see Fig.~\ref{Fig:equator_sigmaB2}), the magnetized flux tube is also smaller in width.
Moreover, the magnetic field lines show a less prominent bend than their prograde counterparts.
Although both cases demonstrate a distinct clockwise shape and rise in height, increasing black hole spin results in a more pronounced bend in the flux tube morphology.
This behavior is related to both the weaker frame-dragging effect $-$and thus slower rotation speeds$-$ at the equatorial plane, as well as the nearly radial topology of the jet funnel that the flux bundle was previously connected to (depicted in orange in Fig. \ref{Fig:flux_tube}).
This effect will become more apparent in the flare trajectories\footnote{Hereafter, the term 'flare trajectory' refers to the measured flux centroid positions derived from the GRRT calculations for a flare model.} discussed in section~\ref{Sec:Flares}.

On the other hand, retrograde simulations display a different behavior.
The newly generated flux tube gains height above the equatorial plane and initially starts winding in the clockwise direction $-$co-aligning with the spin of the black hole$-$, as shown in Fig.~\ref{Fig:flux_tube}.
As the magnetic field lines expand outward, they experience a sudden turn in the opposite direction a few gravitational radii above the equatorial plane.
Following this turning point, the flux tube changes direction and wraps around the magnetized jet funnel in a counterclockwise manner $-$co-aligning with the disk rotation$-$.
Its shape reflects the morphology of the jet funnel it was previously attached to, which is also winding in a counterclockwise manner (see orange lines in Fig.~\ref{Fig:flux_tube}).
This behavior has also been observed in \cite{Porth_2021}, who investigated the motion of orbiting flux bundles in the MAD state. 

In nonnegative spin cases, the flux tube shows a consistent clockwise morphology (see also \cite{Antonopoulou2025}).
In retrograde simulations, however, the magnetic field lines within the active region quickly turn in the counterclockwise direction in response to the black hole's spin.
Surprisingly, this effect is less pronounced in the maximally rotating case.
As shown in Fig.~\ref{Fig:equator_sigmaB2}, the active region significantly increases in size for $a=-15/16$, leading to a more extended and less structured flux tube morphology.
As a result, a few magnetic field lines anchored within the active region do not experience this turning point and continue to rise in a clockwise manner above the disk plane.
However, these lines are a minority, and the general flux tube trend shifts toward a counterclockwise shape (see rightmost panel of Fig.~\ref{Fig:flux_tube}).
Overall, the magnetic field lines comprising the flux tube continuously rise in height and ultimately rotate in the same sense as the accretion disk $-$opposite to the black hole spin$-$ in all of the retrograde simulations investigated.

\subsection{Flare Dynamics} \label{Sec:Flares}
This section focuses on the flare kinematics on the sky plane following a prominent flux eruption.
A constant three-dimensional flux tube morphology derived from the simulation is assumed throughout the event. 
We consider a spherical hot spot with an effective emission radius of $1\,$M that would be formed within the active region on the equatorial plane of the disk. 
Then, we model its trajectory in the sky, assuming it would be ejected along the magnetized flux tube at a relativistic velocity. 

\subsubsection{Prograde spin} \label{subsec:pro}
\begin{figure*}
\sidecaption
\includegraphics[width=5.75cm]{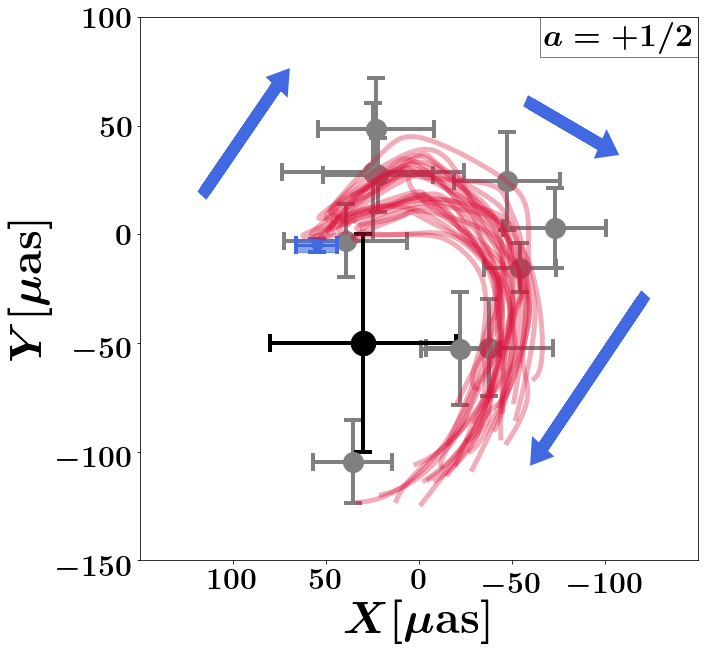}
\hspace{0.0cm}
\includegraphics[width=6.65cm]{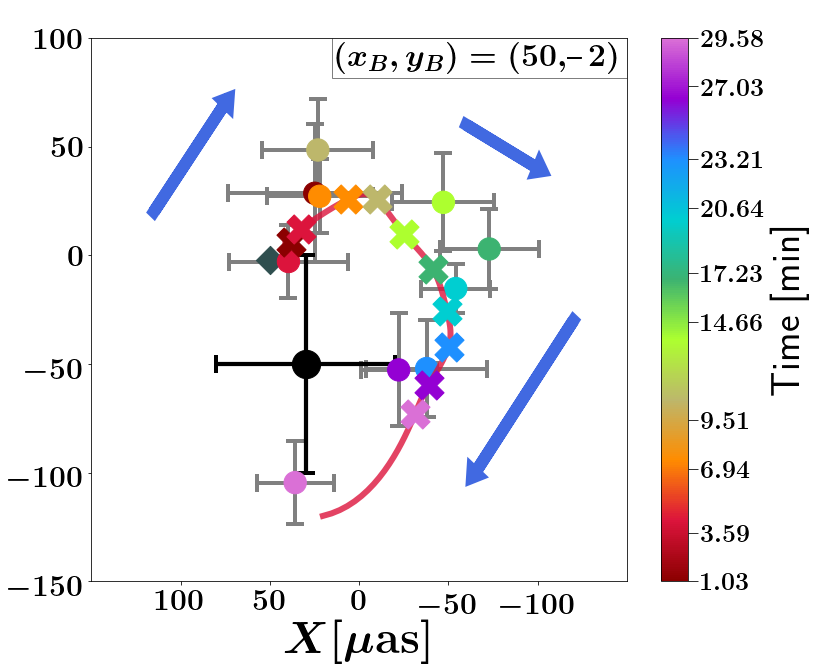} 
\caption{Hot spot trajectories during a prominent flux eruption in the prograde spinning case ($a=+1/2$) overlapped with the observations of July 22 2018 (gray and colorful circles, respectively). The direction of motion is indicated by blue arrows. The position of \sgra in the sky is denoted by a black cross. \textit{Left Panel}: Flare models for hot spot ejection velocities between $0.5c-0.8c$ and foot-point orbital velocities ranging from $0.1u_K-0.3u_K$. The best-fit black hole position, slightly different for each model, is illustrated by a blue rectangle. \textit{Right Panel}: A flare model providing good agreement with the GRAVITY observations. The position of the black hole is depicted by a black diamond.}
\label{Fig:Flare_a+12}
\end{figure*}

Fig.~\ref{Fig:Flare_a+12} shows the flare kinematics for the moderately spinning prograde case, $a_{BH}=+1/2$.
The left panel illustrates several flare models overlapped with the observations of 2018 July 22. 
The hot spot ejection velocity in these models varies between $0.5\,$c to $0.8\,$c, whereas the orbital velocity in the accretion disk ranges from $0.1\,u_K$ to $0.3\,u_K$, based on the simulation.
We readily confirm that the ejected hot spots can balance the counterclockwise drag of the accretion disk and demonstrate a clockwise motion in the sky for a moderate prograde spin.
The flare models present a clear, consistent clockwise trend that is in good agreement with the observed flux centroid positions.

\begin{figure*}[h]
\sidecaption
\includegraphics[width=5.75cm]{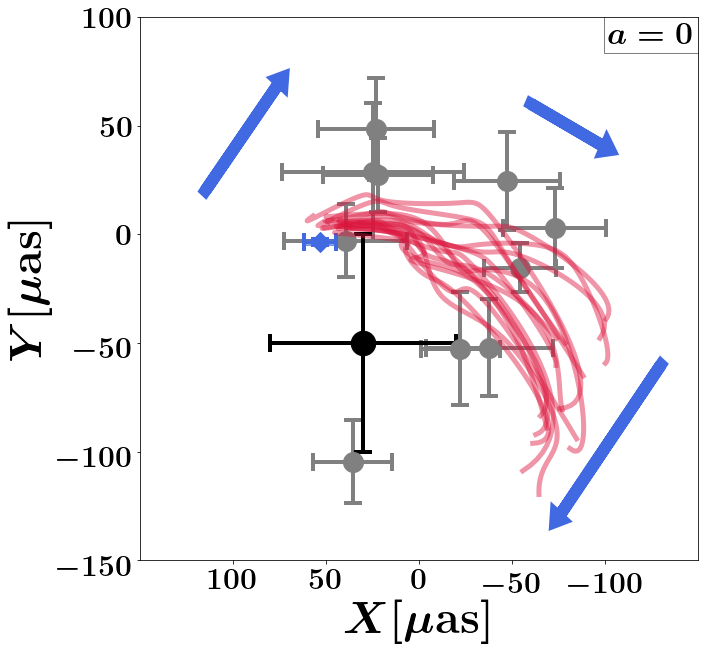}
\hspace{0.0cm}
\includegraphics[width=6.65cm]{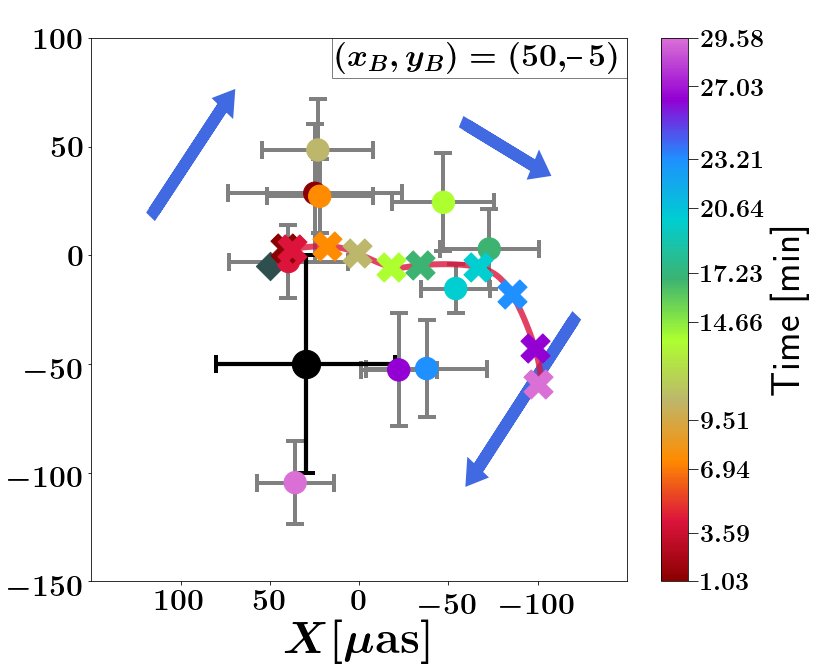}
\caption{Same as Fig.~\ref{Fig:Flare_a+12} for the non-spinning simulation. \textit{Left Panel}: Flare models with ejection velocities of $0.7c-0.8c$ and foot-point orbital velocities between $0.1u_K-0.2u_K$. \textit{Right Panel}: A flare model showing similarities to the GRAVITY observations.}
\label{Fig:Flare_a0}
\end{figure*}
The flare kinematics in the moderately spinning case are also consistent with the maximally rotating case investigated in \cite{Antonopoulou2025}.
The illustrated hot spot trajectories closely reflect the morphology of the magnetized flux tube, which is very similar in both cases (see section~\ref{Sec:Flux_Tube}).
Despite the counter-clockwise drag of the accretion disk, the flare models show a consistent clockwise evolution, passing within the observed error bars of most centroid positions.
The parameter space is also very similar between the prograde simulations, with the moderate spin models being able to account for a significantly sub-Keplerian disk rotation coupled with a relativistic ejection velocity.
In general, flare models with a faster orbital velocity in the accretion disk require a larger ejection velocity, due to the counter-clockwise rotation.
As the hot spot is ejected from the equatorial plane and traverses along the clockwise-shaped flux tube, its foot-point on the disk is dragged in the opposite direction by the accretion flow.
The larger the accretion disk rotation, the faster the hot spot velocity required to balance it.

A good example is given in the right panel of Fig.~\ref{Fig:Flare_a+12}, which depicts a flare model with an ejection velocity of $0.6\,$c and an orbital velocity of $0.1\,u_K$.
This model shows good agreement with the observations of July 22, closely following the flare kinematics and passing within the observed error bars for all but the most distant centroid position.
It is also broadly consistent with the morphology of the best-fitting flare illustrated in \cite{Antonopoulou2025}.
Although the general trend in the moderately spinning flare models is consistent with the observations, many of them still fall short in accounting for the first and final observations of July 22, as shown in Figure~\ref{Fig:Flare_a+12}.

\subsubsection{Non-spinning} \label{subsec:zero}
Fig.~\ref{Fig:Flare_a0} shows the flare kinematics for the non-spinning case, $a_{BH}=0$. 
The left panel depicts the flare models overlapped with the July 22 observations.
We show that a relativistic hot spot ejected along the magnetized flux tube can counteract the drag of the accretion disk and move clockwise in the sky, even for a Schwarzschild black hole.
The ejection velocity in this case varies between $0.7\,c$ and $0.8\,c$, while the orbital velocity on the disk ranges from $0.1\,u_K$ to $0.2\,u_K$, according to the simulation.
The models exhibit a clockwise trend and are similar to the observations of July 22.  

Furthermore, the flare models illustrated in Fig.~\ref{Fig:Flare_a0} demonstrate characteristics similar to those of their prograde spinning counterparts. 
Both the moderately spinning case investigated here (see Fig.~\ref{Fig:Flare_a+12}) and the maximally rotating simulation detailed by \cite{Antonopoulou2025} show a consistent clockwise trend following a prominent flux eruption event.
Nevertheless, it is clear that the hot spot trajectories in the non-spinning case display a smaller clockwise bend and are concentrated in the central region of the observed centroid positions. 
This behavior arises from the underlying magnetic field line topology $-$ eg. a more radial jet funnel; see Fig. \ref{Fig:flux_tube} $-$ and the absence of the frame-dragging effect $-$ leading to slower rotation speeds on the disk plane $-$.  
As a result, the flare models provide scarce coverage of the observed flaring event and fail to reproduce the signature bend in the data set on July 22.

The right panel of Fig.~\ref{Fig:Flare_a0} shows a flare model trajectory from the Schwarzschild simulation plotted with the observations of July 22, 2018.
This model has an ejection velocity of $0.7\,$c and an orbital velocity of $0.2\,u_K$, similar to the prograde flare model depicted in the right panel of Fig.~\ref{Fig:Flare_a+12}.
The illustrated hot spot trajectory exhibits reduced curvature and falls within the observed error bars for only a select few of the centroid positions.
As most of the non spinning models, it shows similarities to the flare kinematics at earlier times, but fails to reproduce the furthest observed centroid positions. 

\begin{figure*}[h]
\sidecaption
\includegraphics[width=5.75cm]{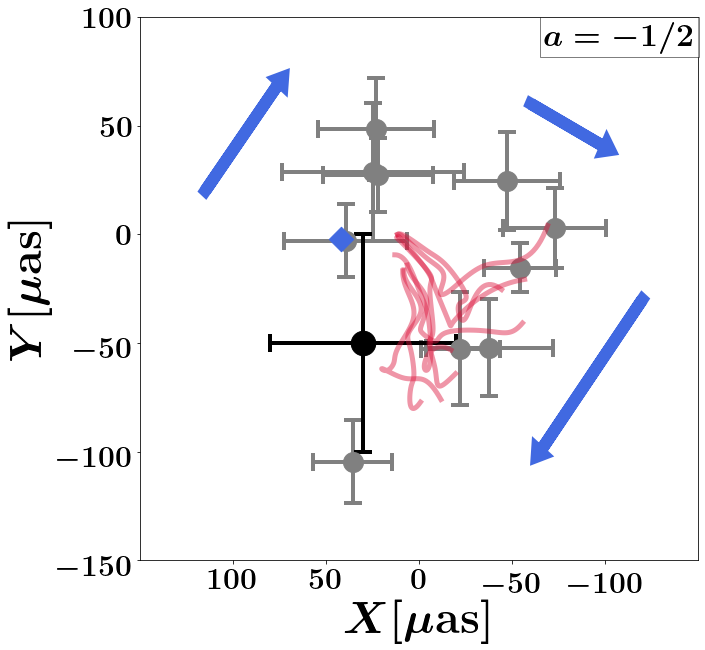}
\hspace{0.0cm}
\includegraphics[width=5.3cm]{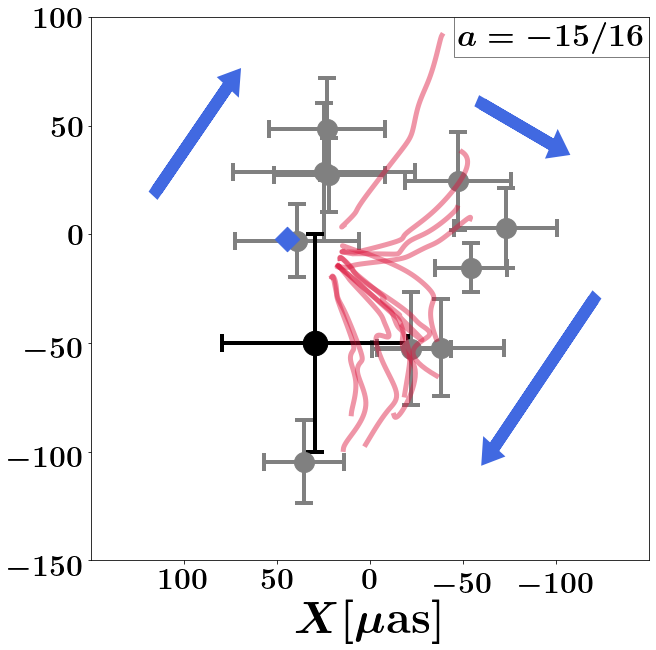} 
\caption{Same as Fig.~\ref{Fig:Flare_a+12} for retrograde simulations featuring a counterclockwise accretion disk. \textit{Left Panel}: Flare models for a moderate spin ($a=-1/2$). Hot spot ejection velocities range between $0.4c-0.5c$ and foot-point orbital velocities from $0.1u_K$ to $0.2u_K$. \textit{Right Panel}: Flare models for a maximally rotating black hole ($a=-15/16$). Hot spot ejection velocities range between $0.3c-0.4c$ and foot-point orbital velocities from $0.1u_K$ to $0.2u_K$.}
\label{Fig:Flare_a-}
\end{figure*}
\subsubsection{Retrograde spin} \label{subsec:retro}
Next, we focus on retrograde spinning black holes with $a_{BH}=-1/2$ and $-15/16$.
In these simulations, the active region formed on the equatorial plane following a flux eruption is significantly larger in size compared to the prograde spinning case (see Fig.~\ref{Fig:equator_sigmaB2}).
The morphology of the magnetized flux tube anchored in this region is thus more extended and less uniform, especially for a maximally rotating black hole.
In both cases, the magnetic field lines show a different behavior from the prograde simulations.
Although they demonstrate a clockwise shape at first, they quickly reach a turning point and wrap in the opposite direction $-$co-aligning with the disk rotation$-$ (see Fig. ~\ref{Fig:flux_tube}).
This behavior is linked to the jet morphology $-$the flux tube was previously connected to$-$ in retrograde simulations.

Fig.~\ref{Fig:Flare_a-} shows a few indicative flare models for a counterclockwise disk rotation. 
A hot spot ejected with a relativistic velocity along the magnetized flux tube still manages to balance the accretion disk rotation in the retrograde spin cases.
However, the resulting flare trajectory closely reflects the underlying flux tube morphology, which, in this case, does not demonstrate a consistent clockwise pattern.
As the hot spot traces the flux tube, its trajectory quickly shifts in the counter-clockwise direction, moving away from the observed centroid positions.
In both cases, the retrograde models either trace a small part of the flare trajectory in the sky before reaching a turning point or move in the completely opposite direction.
During our investigation, we were unable to reproduce the July 22 observations for any combination of model parameters.
As a result, we deduce that retrograde spin models with a counterclockwise accretion disk are not consistent with the observed flare kinematics and provide a poor fit to the astrometric observations of July 22, 2018.

In retrograde simulations, the flux tube shape initially wraps around the jet funnel in the same direction as the black hole is spinning.
After rising a few gravitational radii above the equatorial plane, the magnetic field lines turn and wrap in the direction of the disk's rotation, as discussed in Section \ref{Sec:Flux_Tube}. 
This behavior motivates us to investigate the flare kinematics for a clockwise accretion disk.
In this case, the flux tube morphology co-aligns with the disk rotation and demonstrates a clockwise bend as it rises in height.
Fig. ~\ref{Fig:Flare_a-_clock} shows a few representative flare models for the retrograde simulations in this work.
We readily confirm that flare models featuring a clockwise accretion disk show similarities to the observations of \cite{GRAVITY_2018}. 
The flare trajectories for a spin of $a=-1/2$ are able to reproduce the early-time flare kinematics, however, they do not exhibit the strong clockwise bend necessary to reach the late time observations.
For a maximally rotating black hole, the flare models show an increased clockwise bend and provide a very good fit to the July 22 astrometric observations.
The flare trajectories in this case pass within the observed error bars for most flare positions, as illustrated in the right panel of Fig. ~\ref{Fig:Flare_a-_clock}.
This result serves as a strong metric to differentiate between: (a) prograde and retrograde spinning black holes in our flare model, and (b) accretion disk rotation for a negative black hole spin in the observed flares.

\subsection{Goodness of fit} \label{subsec:fit}
To evaluate the goodness of fit between the flare model trajectory and the \cite{GRAVITY_2018} observations, one needs to take into account two important factors: (a) the deviation of the flare position in the model from the flux centroid observed in the sky, and (b) if the model flare position is within the observed error bars at each timestamp.
To quantify the deviation between model and observation, we employ a reduced $\chi^2$ metric, evaluated as follows:
\begin{equation}
    \label{χ^2}
    \chi_r^2 \,\equiv\, \dfrac{1}{\,2N-N_f\,} \sum_{i\,=\,1}^{N}\left[ \left(\dfrac{X(t_i)-X_i}{\sigma_{X_i}} \right)^2 + \left(\dfrac{Y(t_i)-Y_i}{\sigma_{Y_i}} \right)^2 \,\right] \,,
\end{equation}
where $N$ is the number of observations, $t_i$ is the $i$th observation time, $(X_i\pm\sigma_{X_i}, Y_i\pm\sigma_{Y_i})\,$ is the $i$th observed flux centroid position with its associated error bars, $(X(t_i), Y(t_i))$ is the calculated flux centroid position in the model and $N_f$ is the number of free parameters. 
The flare models have eight free parameters in total: two for the orientation of the observer $(\theta_{obs}, \phi_{obs})$, two for the black hole's location in the sky $(x_{B},\, y_{B})$, two for the hot spot's initial position on the orbital plane $(x_{spot},\,y_{spot})$, one for the accretion disk rotation and one for the hot spot ejection velocity. 
We search for values of $\chi_r^2$ close to unity that indicate a good correlation between the model and observations.
In the process of finding the best-fit value for $\chi_r^2$, we follow a procedure for determining the black hole position in the sky, so that as many flare positions as possible fall within the corresponding error bars. 

\begin{figure*}[h]
\sidecaption
\includegraphics[width=5.75cm]{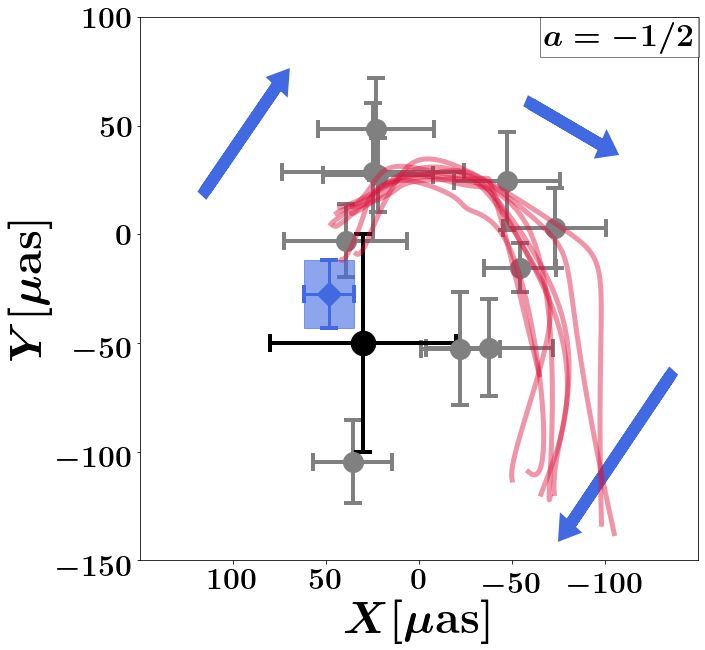}
\hspace{0.0cm}
\includegraphics[width=5.3cm]{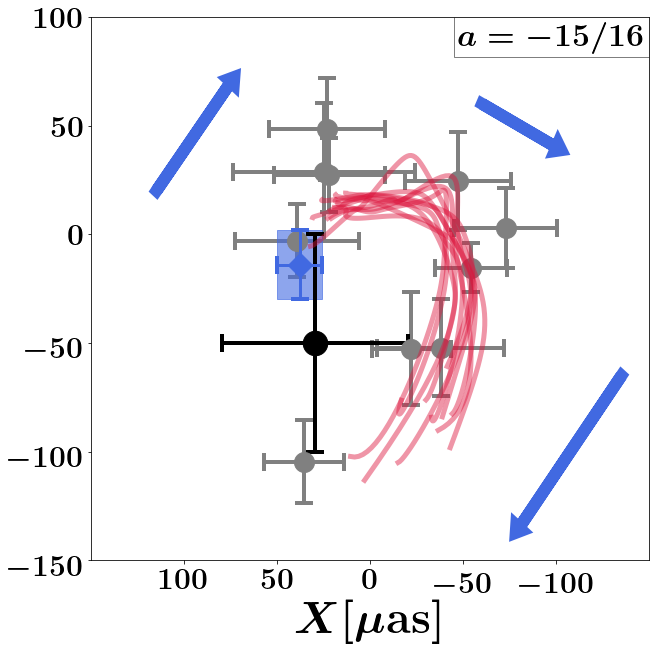} 
\caption{Same as Fig.~\ref{Fig:Flare_a+12} for retrograde simulations featuring a clockwise accretion disk. \textit{Left Panel}: Flare models for a moderate spin ($a=-1/2$). Hot spot ejection velocities range between $0.5c-0.7c$ and foot-point orbital velocities from $0.6u_K$ to $0.8u_K$. \textit{Right Panel}: Flare models for a maximally rotating black hole ($a=-15/16$). Hot spot ejection velocities range between $0.4c-0.7c$ and foot-point orbital velocities from $0.6u_K$ to $0.9u_K$.}
\label{Fig:Flare_a-_clock}
\end{figure*}

Based on the formulation outlined above, we calculate the $\chi_r^2$ value for the flare models depicted in Figures ~\ref{Fig:Flare_a+12} through ~\ref{Fig:Flare_a-_clock}.
In order to qualitatively evaluate the goodness of fit in each case, we select a conservative threshold of $\chi_{r,thres}^2=2.5$ and calculate the fraction of flare models with a $\chi_r^2$ below the adopted threshold for each black hole spin.
Table \ref{Table:Fit_x2} shows the pass fraction and average $\chi_r^2$ for all models investigated in this work, including the maximally rotating prograde models presented in \cite{Antonopoulou2025} for completion.
It is readily confirmed that Kerr black holes provide a much better fit to the observed flare kinematics.
As evident in Fig. ~\ref{Fig:Flare_a+12}, prograde flare trajectories pass within most of the observed error bars and thus result in a $\bar{\chi}_r^2$ value closer to unity compared to their nonspinning counterparts.
This is mainly attributed to the less pronounced clockwise bend in the flux tube morphology within the nonspinning simulation.
The Schwarzschild pass fraction for the illustrated flare models is just below $20\%$, whereas the prograde models are much more favored with a pass fraction of $69\%$.
In addition, both moderately and maximally rotating prograde models are highly favored in this analysis, with the counterclockwise models presented in Fig. 3 of \cite{Antonopoulou2025} reaching a pass fraction of $95\%$ and a $\bar{\chi}_r^2\simeq1.34$.

On the other hand, counterclockwise retrograde models fail to reproduce the observed flare kinematics in the sky (see Fig. ~\ref{Fig:Flare_a-}).
Instead, retrograde models provide a better fit to the July 22 flare trajectory when the accretion disk is rotating in the clockwise direction.
As evident in Fig. ~\ref{Fig:Flare_a-_clock}, flare models for a clockwise disk rotation show a clockwise motion in the sky and an increasing bend in the trajectory for larger (absolute) spin magnitude.
For $\,\abs{a}=1/2$, prograde models are favored due to the enhanced clockwise bend in the flare trajectories, with a pass fraction reaching $69\%$.
However, for a maximally spinning black hole, both prograde and retrograde models are plausible, with pass fractions above $90\%$\footnote{The number of flare models investigated differs between each case. A representative sample of flare trajectories has been selected for each black hole spin and accretion disk rotation in this work, ranging between $8-26$ models depending on the configuration. The maximally rotating case investigated in \cite{Antonopoulou2025} spans approximately $20$ models in the counterclockwise regime, and only $5$ models in the clockwise regime, respectively.\label{fn:models}}.
The flare trajectories show a distinct clockwise trend and pass within most of the observed error bars of the July 22 flare in both cases.
Retrograde models are more favored in this configuration, as their $\bar{\chi}_r^2$ is closer to unity.

In summary, our analysis favors a Kerr black hole with a preference towards larger spin magnitude.
A counterclockwise disk scenario supports a prograde black hole with $a\geq1/2$.
Retrograde configurations are incompatible with the observed flare kinematics in the Galactic center in this picture.
Conversely, a clockwise accretion disk rotation favors a maximally rotating black hole with $\abs{a}=15/16$.
Spin orientation is inconclusive for the clockwise scenario. 
Retrograde models provide an average $\chi_r^2$ closer to unity, and are more favored by the current analysis. 
However, prograde models are also able to account for the observed flare kinematics, despite the limited sample size presented in \cite{Antonopoulou2025}.
\par
While the present study investigates a limited sample in spin, we expect intermediate values to demonstrate a behavior in-between the investigated cases.
In particular, small values of spin (eg. close to zero) are expected to provide a poor fit to the observed flare kinematics $-$see Fig. \ref{Fig:Flare_a0}$-$.
Increasing spin values would progressively show a stronger bend in the underlying flux tube morphology and flare trajectory in the sky $-$see Fig. \ref{Fig:Flare_a+12}$-$ that would resemble the observations of \cite{GRAVITY_2018}.

\begin{table}[h] 
\centering
\caption{Goodness of fit evaluation.}                   
\begin{tabular}{c c c c}  
\hline\hline 
Spin & Disk rotation & Pass fraction & $\bar{\chi}_r^2$ \\
\hline
$+15/16$\tablefootnote{The maximally rotating prograde flare models are illustrated in \cite{Antonopoulou2025}. We refer the reader to Fig. 3 and G.1. in the corresponding paper.} & counterclockwise & $0.95$ & $1.34$ \\
$ $ & clockwise & $0.6$ & $2.6$ \\
\hline
$+1/2$ & counterclockwise & $0.69$ & $2.19$ \\
\hline
$0$ & counterclockwise & $0.18$ & $4.8$ \\
\hline
$-1/2$ & counterclockwise & $0.0$ & $7.02$\\
$ $ & clockwise & $0.25$ & $3.79$\\
\hline
$-15/16$ & counterclockwise & $0.0$ & $7.11$\\
$ $ & clockwise & $0.91$ & $1.94$\\
\hline   
\end{tabular}
\tablefoot{The first column denotes the spin of the black hole, and the second column specifies the accretion disk rotation. The third column shows the pass fraction for each configuration ($N_{pass}/N_{models}$\footref{fn:models}), assuming a conservative threshold of $\chi^2_r=2.5$. The fourth column depicts the average $\chi_r^2$ for all models.}      
\label{Table:Fit_x2} 
\end{table}
\section{Conclusions} \label{Sec:Conclusions}
In summary, this work highlights the connection between quasi-periodic flux eruption events and the observed flare kinematics in the Galactic center. We have analyzed a series of three-dimensional GRMHD simulations across a range of black hole spins, including our previous work \cite{Antonopoulou2025}. Our investigation includes two prograde and two retrograde cases, for an average and maximal black hole spin, as well as the non-rotating Schwarzschild case. 

In each simulation, a prominent flux eruption is selected from the time series of the normalized magnetic flux on the horizon, the evolution of the active region on the equatorial plane is monitored throughout the event, and the topology of the newly generated flux tube (whose footpoints are anchored in this region) is recovered. 
The energetic flares are modeled via a spherical hot spot ejected from the disk plane and traversing along the magnetized flux tube at a relativistic velocity. 
We choose a face-on observation angle and trace the kinematics of our flare model and its dependence on the spin of the black hole. 
The flares models are compared to the astrometric observations of \cite{GRAVITY_2018} and the goodness of fit is evaluated via a reduced $\chi^2$ metric, which quantifies their deviation at each time stamp.
A conservative threshold of $\chi^2_{r,thres}=2.5$ is then selected in order to determine the fraction of passing models per black hole spin configuration.
 
This study uncovers a unique dependence between flares in the vicinity of \sgra in relation to black hole spin. 
The flare kinematics in the sky strongly depend on the underlying flux tube morphology, which is generated close to the supermassive black hole and is thus related to its spin. 
For example, the Schwarzschild flare models show poor agreement with the \cite{GRAVITY_2018} astrometric observations in comparison to their Kerr counterparts, since the magnetic field lines' topology is more vertical and demonstrates only a small clockwise bend.
They have an $\chi_r^2\sim4.8$ that is far from unity, reflecting the poor fit the Schwarzschild models provide to the observed flare kinematics.
In the nonspinning case, we do not expect a change in disk rotation to significantly affect the flare trajectory in the sky.

In the Kerr framework, the flare kinematics are also tied to the accretion disk rotation. 
In prograde simulations, the generated flux tube wraps around the jet funnel in the opposite direction to the black hole spin, thus opposite to the general accretion flow, and demonstrates a stronger bend in the respective direction for larger spin magnitude. 
The resulting flare trajectories show good agreement with the observed flare kinematics for both a maximally and a moderately rotating black hole featuring a counterclockwise accretion disk.
As shown in \cite{Antonopoulou2025}, prograde models exhibit similar characteristics for a clockwise rotating disk, although these flare models require nearly Keplerian orbital velocities in the equatorial plane to reproduce the July 22 observations. 

In contrast, the flux tube morphology in retrograde simulations changes drastically.
The flux tube quickly reaches a turning point and wraps around the jet funnel in the opposite direction to the black hole spin.
This means that for a counterclockwise accretion flow, the magnetic field lines in the flux bundle show a counterclockwise bend, and vice versa.
The ejected hot spot trajectories considered trace the underlying flux tube and are thus inconsistent with the observed clockwise flare of July 22 for a counterclockwise disk.
However, our flare models show good agreement between the observed flare kinematics and a clockwise disk rotation, especially for a maximally rotating black hole.
In this case, flare trajectories demonstrate a strong clockwise bend and are able to recover the observed kinematics of the investigated flare, showing an $\bar{\chi}_r^2\sim1.94$.   

All investigated MAD simulations show a flux tube that winds in the opposite direction to the black hole's spin, reflecting the morphology of the magnetized jet funnel. 
Therefore, our results provide a promising framework to constrain spin magnitude and orientation from the observed flare kinematics.
The present analysis favors Kerr black holes featuring a large spin magnitude, with pass fractions surpassing $90\%$ for $\abs{a}=15/16$.
Specifically, assuming a counterclockwise accretion disk, our flare models indicate a prograde black hole spin of $a\geq1/2$.
In a clockwise disk scenario, both prograde and retrograde models provide a good fit to the observed flare kinematics of July 22, although the retrograde models provide $\chi_r^2$ values closer to unity.
We expect that assuming an intermediate value of spin in the GRMHD simulation setup would result in a behavior in-between the investigated cases.
\par
In conclusion, this study shows that flare models powered via strong flux eruption events in MAD simulations are good candidates for reproducing the astrometric observations of \cite{GRAVITY_2018} in the Galactic center.
The models are able to capture the observed flare kinematics $-$that show a consistent clockwise motion in the sky \citep{GRAVITY_2018,GRAVITY_2020,GRAVITY_2023}$-$ for a range of black hole spin.
Our analysis favors Kerr black holes with a spin of $\abs{a}=15/16$, although milder spin configurations are not entirely excluded.
These results are consistent with the horizon-scale images of \sgra and the multiwavelength characteristics of the source \citep{EHTV2022}, as well as simultaneous analysis on its polarimetric properties across different scales \citep{yfantis_2023,EHTVIII2024}.
Future improvements to the EHT array (eg. the Black Hole Explorer mission \citep{Johnson_2024}) and the GRAVITY instrument of the Very Large Telescope will enhance our accuracy in recovering orbital motion in the Galactic center and enforce stricter constraints on the spin of \sgra \citep{Yfantis2026,Conroy2026}.

\section{Discussion} \label{sec:Discussion}

A significant drawback of MAD models is that they produce a strong jet that comes at odds with the absence of a strong, extended jet structure for \sgra. \cite{EHTVIII2024} has calculated an outflow power of $4\times10^{38}\,$erg s$^{-1}$ for one of the MAD models in the best-bet region, assuming a mass accretion rate of $5\times10^9\,M_{\odot}$yr$^{-1}$. The resulting jet efficiency of $\sim150\%$ in this model, powered by the \cite{BLandZnaj} mechanism, is not large enough to produce global effects on the evolution of the Milkyway \citep{Su2021}. Although there has been evidence of extended structures and active winds originating from the Galactic center in recent years \citep{Su_2010,Gorski_2026}, suggesting strong activity close to the supermassive black hole, \sgra does not show an extended observed jet, in contrast to MAD simulations. As a result, the flare model underlying this work could be well-suited to describe flaring events in the vicinity of supermassive black hole systems exhibiting a strong jet, such as \m87.
\par
Assuming more realistic initial conditions for the system, such as a low-angular momentum disk, a more Bondi-like accretion flow or larger scale stellar winds feeding the supermassive black hole would produce a weaker jet and reduced orbital flux tube motion \citep{Ressler_2020,Ressler_2021,lalakos2025,cho2025,Galishnikova_2025}.
Our flare model relies on the generation of magnetized flux tubes, disconnected from the jet sheath close to the supermassive black hole and orbiting at the local velocity of the accretion flow, and the formation of energetic hot spots, powered by magnetic reconnection events on the equatorial plane.
The ejection of hot spots along the magnetized flux tube can be attributed to the generation of strong non-ideal electric fields during the violent flux eruption events in the MAD state. 
We note that the inclusion of resistivity is essential to capture these effects \citep{Ripperda2019,Nathanail2025}, that are not present in the ideal GRMHD simulation regime.
Investigating a weaker version of these effects in a more realistic simulation setup constitutes a necessary future step of the current analysis.
Nevertheless, the selection of a MAD is a well-motivated initial study, as these models have been thoroughly compared against numerous observational constraints of \sgra and are capable of reproducing the majority of the intensity and polarimetric constraints of the source \citep{EHTV2022,EHTVIII2024}.
\par
Furthermore, the dissipation time scale of magnetized flux tubes depends on the grid resolution within the GRMHD simulation \citep{Salas_2024,Ripperda_2022}.
Specifically, it has been shown that the resolution utilized in this work is prone to diffusion of the flux tube beyond one orbital rotation \citep{Porth_2021,Salas_2024}.
At higher resolution, the flux tubes tend to survive for a longer duration and move towards larger radii in the accretion flow.
Nevertheless, the emitting hot spot considered in this model is expected to cool down due to adiabatic expansion and the synchrotron cooling timescale prior to flux tube  dissipation due to the limited resolution. 
In addition, our flare model investigates less than a full orbit of the flux tube, and it is therefore not affected by numerical diffusion. 
As a result, the resolution of the GRMHD simulations underlying this work is sufficient to capture the flare kinematics in the vicinity of \sgra.
\par
Future work includes the investigation of the polarized signature of the flare models outlined in this work and in \cite{Antonopoulou2025}.
Recent flare observations demonstrate both a relativistic clockwise trajectory in the plane of the sky and a matching evolution in polarization angle over the duration of an hour \citep{GRAVITY_2023}, with several flares being accurately detected solely in the Q-U plane.
Initial investigations hint at a clockwise Q-U loop in the sky that demonstrates a similar behavior to the observed flares in the vicinity of \sgra.
Furthermore, a combined study of the lightcurve and spectral properties of the observed flares in the Galactic center is deemed necessary.
In recent years, James Webb Space Telescope has reported the evolution of the spectral index in the mid-infrared \citep{vonFellenberg_2025}, which should also be smoothly connected to the X-ray regime.
These observations place strong constraints on the eDF and the mechanism to accelerate emitting particles close to the supermassive black hole. 
Although this study considers a fixed power-law index, we do not expect this choice to have a strong impact on the observed flare kinematics in the sky plane.
A detailed investigation of the required eDF and a combined study of the astrometric and polarimetric properties, light curve, and spectral index evolution in the detected Galactic center flares is left for future work.

\begin{acknowledgements}
EA acknowledges financial support by the A.G. Leventis foundation. 
AN \& FZ are supported by the Hellenic Foundation for Research and Innovation (ELIDEK) under Grant No 23698. 
EA and AN, from the OSCARS-funded project MEXANE, acknowledge the OSCARS project, which has received funding from the European Commission's Horizon Europe Research and Innovation programme under grant agreement No. 101129751.
YM is supported by the National Key R\&D Program of China (grant No. 2023YFE0101200), the National Natural Science Foundation of China (grant Nos. 12273022 and 12511540053), and the Shanghai municipality orientation program of basic research for international scientists (grant No. 22JC1410600).
This work was supported by computational time granted from the National Infrastructures for Research and Technology S.A. (GRNET S.A.) in the National HPC facility - ARIS - under project BHIM.  

\end{acknowledgements}

\nocite{*}
\bibliographystyle{aa} 
\bibliography{rev1_sgra_flares_spin} 

\begin{appendix} 

\section{Numerical setup} \label{Append:Numerical_Setup}
The GRMHD simulations investigated in this work are performed in three spatial dimensions with the Black Hole Accretion Code (BHAC) \citep{Porth_2017,Olivares_2019}\footnote{\, https://www.bhac.science}, using modified Kerr-Schild coordinates \citep{McKinney_2004}.
The base resolution is $384\times 192\times 192$ grid cells in the $r$, $\theta$, and $\phi$ directions, with additional adaptive mesh refinement being employed. 
The simulations are initialized with a \cite{Fishborne_Moncrief_1976} torus in hydrodynamic equilibrium orbiting an uncharged black hole of dimensionless spin parameter ranging between $[-15/16,+1/2]$
The inner edge of the disk initially rests at $r_{in}= 20\,M$\footnote{\,In units where $c=G=1$, a length unit is equal to the black hole mass, $r_g=M$.} and the density maximum is located at $r_{max} = 40\,M$. 
The equation of state is ideal, with an adiabatic index of $\hat{\gamma}= 4/3$.
The magnetic field in the disk is initialized as a nested loop described by the vector potential
\begin{equation}
    \label{vector_potential}
    A_{\phi} = max \left( \left( \dfrac{\rho}{\rho_{max}} \, \left(\dfrac{r}{r_{in}}\right)^3 \, \sin^3\theta\,\exp{\left(\dfrac{-r}{400}\right)} \right) - 0.2, 0 \right) \,,
\end{equation}
where $\rho_{max}$ is the maximum rest-mass density in the torus.
\begin{figure}
\centering
\includegraphics[width=0.5\textwidth]{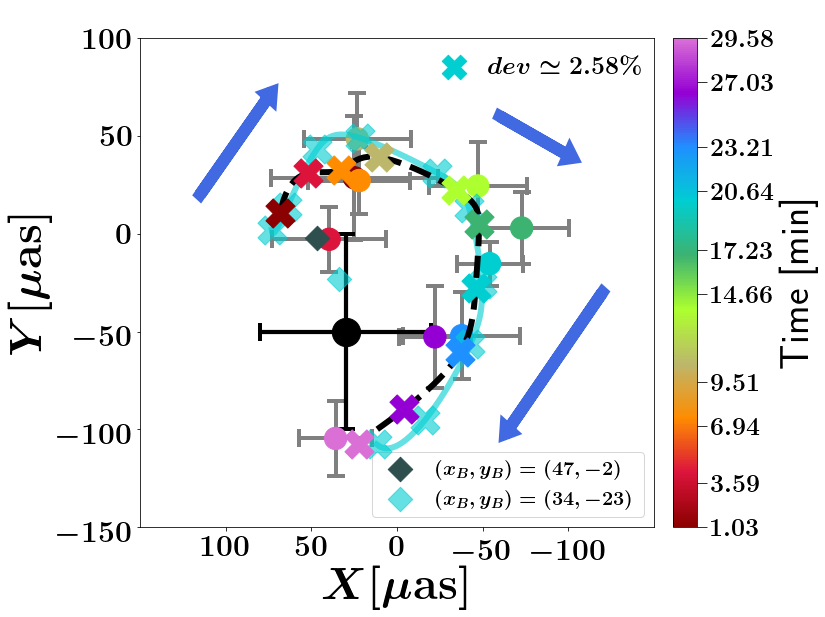}
\caption{Comparison between the resulting flare trajectory assuming a fixed (cyan line) and an interpolated flux tube morphology (dashed black line). The black hole position for each model is indicated by a diamond of the corresponding color. The flare positions for a fixed flux tube morphology are depicted by cyan crosses (see Fig. 3 in \cite{Antonopoulou2025}). In the interpolated case, the flare positions are color-coordinated with the corresponding observational timestamps. The relative percentile deviation between the flare positions at each timestamp is denoted on the top right legend.}
\label{Fig:interp_traj}
\end{figure}
\begin{figure*}[h!]
\centering
\includegraphics[width=1\textwidth]{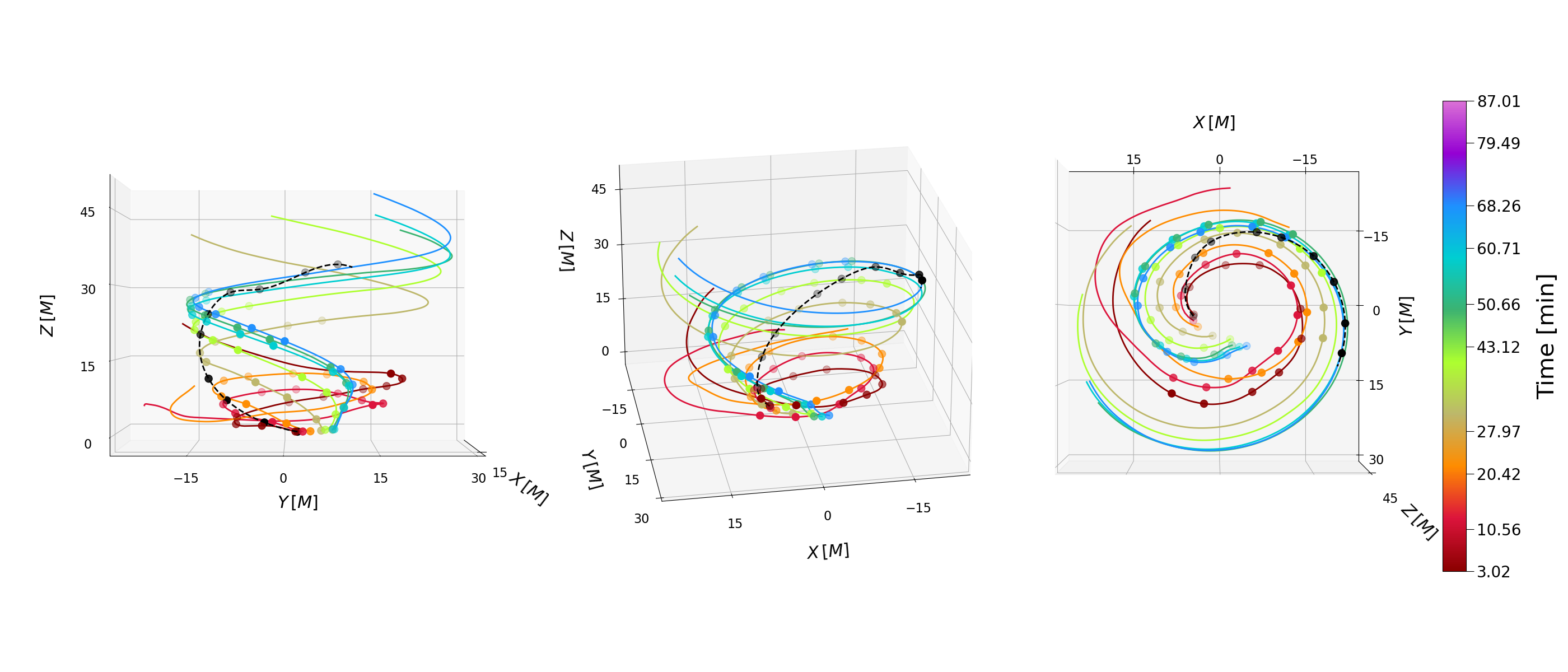}
\vspace{-1cm}
\caption{Three dimensional view of the magnetized flux tube investigated in \cite{Antonopoulou2025}. The evolution of the indicative magnetic field line is illustrated in color, in accordance with the observed flare positions. The interpolated magnetic field line is depicted by a dashed black line. The position of the hot spot as it traverses each magnetic field line is illustrated in colored circles.}
\label{Fig:Flux_tube_evolution}
\end{figure*}

\begin{table*}[h] 
\centering
\caption{GRMHD simulation parameters.}
\begin{tabular}{c c c c c c c c}  
\hline\hline 
Model & a & Resolution & Duration $(M)$ & $\hat{\gamma}$ & $r_{in} (M)$ & $r_{max} (M)$ & $r_{out} (M)$  \\
\hline
MAD & $0,\pm1/2,-15/16$ & $384\times 192\times 192$ & 15,000 & 4/3 & 20 & 40 & 2500 \\ 
\hline   
\end{tabular}
 \tablefoot{Model specifies the accretion disk model, $a$ is the dimensionless black hole spin, Resolution denotes the number of grid zones along each direction as $N_r\times N_{\theta}\times N_{\phi}$, Duration denotes the total duration of the simulation, $\hat{\gamma}$ is the fluid element adiabatic index, $r_{in}$ and $r_{max}$ are the inner and pressure maximum radii of the FM torus, and $r_{out}$ is the outer radial boundary of the simulation domain.}       
\label{Table:GRMHD_Setup} 
\end{table*}
The simulation parameters are summarized in Table \ref{Table:GRMHD_Setup}.
Under these initial conditions, the GRMHD simulations evolve according to the well-investigated MAD regime \citep{EHT_M87_V,wong2021,Mizuno2021,Fromm_2022,narayan2022,EHTV2022,zhang2024}.
MAD simulations are characterized by frequent flux eruption events, due to the saturation of magnetic flux on the event horizon, thus temporarily halting accretion (see discussion in Section \ref{Sec:Methodology}).
This renders them ideal candidates for the generation of magnetized flux tubes, strong magnetic reconnection events, and prominent flare emission.
In contrast, SANE disks do not exhibit these violent flux eruptions \citep{CruzOsorio_2022,Fromm_2022}, and are thus excluded from the present study.
\par
Radiative transfer calculations are performed in Boyer-Lindquist coordinates with the Black Hole Imaging code \citep{Antonopoulou_2024}.
A Kerr black hole rests at the center of the coordinate system, and a grid of $600\times600$ photons is initialized at a distance of $500M$.
The observer inclination is face-on $-i=5\degree$ to eliminate issues on the polar axis$-$.
We employ a fixed flux tube morphology for each simulation and model the ejected hot spot trajectory as it traverses the flux tube at a fraction of the speed of light.
Photons are initialized on the observer's grid and ray traced until they reach the hot spot's sphere of emission ($\sim1\,$M) at any given time.
The radiative transfer calculations consider a nonthermal electron distribution described by a power-law \citep{Pandya_2016}. 
Our calculations do not include synchrotron absorption as a first order approximation.

\section{Flux tube evolution} \label{Apped:Fluxtube_evol}

The modeling framework outlining this work utilizes a fixed flux tube morphology, selected a few time units following the onset of the flux eruption. 
This approach relies on the primary assumption that the flux tube's macroscopic shape quickly stabilizes post-eruption and exhibits negligible structural evolution thereafter\footnote{We note that high resolution simulations show that Rayleigh-Taylor instability fingers are able to protrude, de-magnetize, and even break apart the flux tube morphology in certain scenarios. Nevertheless, assuming a stable macroscopic shape with negligible substructure evolution is a reasonable approach in the current investigation.}.
This appendix evaluates the robustness of our assumption, focusing on the maximally rotating prograde simulation presented in \cite{Antonopoulou2025}.

We derive the three dimensional structure of the magnetic field lines comprising the newly generated flux tube (see Fig. 2 in \cite{Antonopoulou2025}) for a total of eight snapshots within the simulation, corresponding to the first eight observational timestamps.
Fig. ~\ref{Fig:Flux_tube_evolution} illustrates an indicative magnetic field line from the center of the flux tube, color-coordinated with the corresponding timestamp.
In other words, this figure showcases the evolution of the magnetized flux tube during the flux eruption event\footnote{We refer the reader to \cite{Nalewajko_2026} for a supplementary detailed analysis of the magnetic field lines echoing a magnetic flux eruption.}.
Initially, the morphology of the magnetic field lines is confined close to the equatorial plane (red to orange lines).
As the flux eruption advances, the flux tube progressively rises in height and expands outward (yellow lines), until it reaches a relatively stable morphology at later times (green and blue lines). 
Throughout its evolution, the flux tube demonstrates a consistent clockwise shape, enhanced by the drag of the accretion disk on the equatorial plane.

We consider a fixed flux tube shape throughout the event, eg. one of the blue magnetic lines illustrated in Fig. ~\ref{Fig:Flux_tube_evolution}.
An alternative approach would be to interpolate between the flux tube morphology at each simulation timestamp. 
To achieve this, we assume that the hot spot is initially moving on flux tube A at time $t_A$ $-$eg. red line at $t=10.56\,$min$-$ and will be moving on flux tube B at a consecutive time $t_B$ $-$orange line $t=20.42\,$min$-$.
The interpolated magnetic field line is depicted by a dashed black curve in Fig. ~\ref{Fig:Flux_tube_evolution}.
It shows a progressive rise in height, closely reflecting the corresponding flux tube at each timestamp, as well as a consistent clockwise bend, in agreement with the overall morphology.
At the initial stages of the flux eruption, the interpolated line continuously opens up, following the flux tube evolution, until it stabilizes at the opening angle of the final simulation steps (green and blue colored lines).

We model the motion of a hot spot that is assumed to traverse the magnetic field lines with an ejection velocity of $0.8\,$c illustrated in Fig. ~\ref{Fig:Flux_tube_evolution}.
The hot spot positions for the observational timestamps of the July 22 flare are depicted by colorful circles on the corresponding magnetic field line.
GRRT calculations are performed for an indicative stable flux tube morphology near the end of the flux eruption (cyan curve) and the interpolated magnetic field line depicted in black.
The resulting flare trajectories are illustrated in Fig. ~\ref{Fig:interp_traj}, depicted in cyan for a constant flux tube morphology and in dashed black for the interpolated field line. 
Note that the cyan trajectory corresponds to one of the best-fitting models presented in \cite{Antonopoulou2025} (see right panel of Fig. 3).
The black hole position for each model is depicted by a colored diamond and denoted on the bottom right legend.
The values on the top right represent the mean spatial deviation between the models, normalized by the diagonal scale of the field of view.

The cyan flare trajectory has a very similar shape to the interpolated trajectory in black.
The trajectories show larger deviations at the early stages of the flux eruption, when the flux tube is progressively expanding outward and rising in height. 
However, they demonstrate very similar behavior at later times after the flux tube morphology has stabilized.
Assuming the same black hole position, the mean normalized spatial deviation across all individual flare positions in the two models is equal to $6.1\%$.
This deviation decreases even more if one shifts the black hole position in the interpolated model, reaching a minimum of $2.58\%$.
As illustrated in Fig. ~\ref{Fig:interp_traj}, the best fit flare model presented in \cite{Antonopoulou2025} is in very good agreement with the interpolated model investigated here.
As a result, we confirm that employing a fixed flux tube shape, selected a few time units after the onset of the flux eruption when it has stabilized in shape, is a reasonable approach in modeling the flare kinematics.

\end{appendix}
\end{document}